\documentclass[aps,10pt,floatfix,nofootinbib,twocolumn]{revtex4-1}
\usepackage{graphicx,amsmath,amssymb}
\usepackage[usenames,dvipsnames]{color}
\usepackage{ulem}

\newcommand{\beq}{\begin{equation}}
\newcommand{\eeq}{\end{equation}}
\newcommand{\bea}{\begin{eqnarray}}
\newcommand{\eea}{\end{eqnarray}}

\newcommand{\kb}{ {k_{\rm B}} }
\newcommand{\kT}{ \kb T }

\newcommand{\ahum}[1]{``#1''}
\newcommand{\eq}[1]{eq.~(\ref{#1})}

\newcommand{\fig}[1]{fig.~\ref{#1}}
\newcommand{\Fig}[1]{Fig.~\ref{#1}}

\newcommand{\tab}[1]{Table~\ref{#1}}

\begin{document}
\title{Membrane lateral structure: The influence of immobilized particles on domain size}

\author{Timo Fischer}
\affiliation{Georg-August-Universit\"at 
G\"ottingen, Institute of Theoretical Physics, Friedrich-Hund-Platz~1, D-37077 G\"ottingen, Germany}

\author{H. Jelger Risselada}  
\affiliation{Max Planck Institute for Biophysical Chemistry,
Department of Theoretical and Computational Biophysics, Am Fa{\ss}berg 11,
D-37077 G\"ottingen, Germany.}

\author{Richard L.~C.~Vink}
\affiliation{Georg-August-Universit\"at 
G\"ottingen, Institute of Theoretical Physics, Friedrich-Hund-Platz~1, D-37077 G\"ottingen, Germany}

\begin{abstract}
In experiments on model membranes, a formation of large domains of different 
lipid composition is readily observed. However, no such phase separation is 
observed in the membranes of intact cells. Instead, a structure of small 
transient inhomogeneities called lipid rafts are expected in these systems. One 
of the numerous attempts to explain small domains refers to the coupling of the 
membrane to its surroundings, which leads to the immobilization of some of the 
membrane molecules. These immobilized molecules then act as static obstacles for 
the remaining mobile ones. We present detailed Molecular Dynamics simulations 
demonstrating that this can indeed account for small domains. This confirms 
previous Monte Carlo studies based on simplified models. Furthermore, by 
directly comparing domain structures obtained using Molecular Dynamics to Monte 
Carlo simulations of the Ising model, we demonstrate that domain formation in 
the presence of obstacles is remarkably insensitive to the details of the 
molecular interactions.
\end{abstract}

\pacs{65.20.De, 87.16.dt, 64.60.De, 64.70.Ja, 61.20.Ja}

\maketitle

%
%
\section{Introduction}

Membranes are two-dimensional (2D) fluid environments, consisting of many 
different types of lipids and proteins \cite{citeulike:1595800}. The membrane 
constituents are not arranged randomly, but are expected to be spatially 
organized into domains of different size, composition, and dynamics 
\cite{citeulike:5470515, citeulike:4308274, citeulike:4312704, 
citeulike:3042594, citeulike:6562069}. In biological membranes this is important 
because it links to key processes in cells, such as signaling, endocytosis, and 
adhesion \cite{citeulike:6505558, citeulike:6499105}. In artificial membranes, 
domain formation is relevant for applications, ranging from photolithographic 
patterning, spatial addressing, microcontact printing, and microfluidic 
patterning \cite{citeulike:6528926, citeulike:6549481}. To identify the factors 
that control domain formation, and to understand their underlying physical 
mechanisms, is therefore of key practical importance.

One challenge is to account for a heterogeneous domain structure in thermal 
equilibrium \cite{citeulike:5470515}. Due to line tension, one would naively 
expect a multi-domain structure to be unstable as the free energy would be 
reduced by domain coalescence. Indeed, in model membranes, this is precisely 
what is observed \cite{citeulike:3850881, Veatch20033074}. At high temperature, 
these systems are in a mixed state, but they phase separate into 
(macroscopically large) liquid-ordered and liquid-disordered domains at low 
temperature. At the temperature where the phase separation begins to occur (the 
so-called critical temperature), the transition passes through a critical point 
\cite{citeulike:3850881, citeulike:3367620, citeulike:3850871}. At the critical 
point, the membrane is highly heterogeneous, featuring domains of all sizes. 
Hence, at least in model membranes, a non-trivial heterogeneous domain structure 
arises only near the critical point. In line with this, it has been proposed 
that critical behavior might also explain a heterogeneous domain structure in 
biological membranes~\cite{citeulike:3852117}. We note that critical behavior is 
not the only feasible mechanism. Also the coupling between the elastic 
properties of the membrane and its composition has been identified to be a key 
factor affecting domain shape and size \cite{citeulike:8893054, schick:2011, 
citeulike:5045047, citeulike:9473614, citeulike:2802537, citeulike:4025142}. In 
addition, the presence of hybrid lipids that collect at the interface between 
liquid-ordered and liquid-disordered domains could also induce a heterogeneous 
domain structure \cite{citeulike:10418449}.

\subsection{Particle immobilization in membranes}

\begin{figure*}
\begin{center}
\includegraphics[width=18cm]{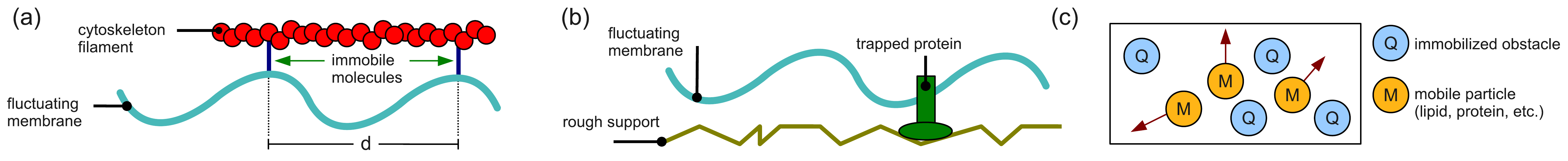}
\caption{\label{fig:qd} Schematic sketches (not to scale) showing possible 
origins of quenched disorder in membranes. (a) In biological membranes, quenched 
disorder may arise from the anchoring of proteins and other molecules to the 
cytoskeleton network. Such anchored proteins are effectively immobilized. The 
typical distance between immobilized particles can become very small $d \sim 
2-9$~nm~\cite{citeulike:6859704}. (b) An analogous {\it in vitro} example is 
provided by supported membranes. In this case, particle immobilization may arise 
from surface roughness or, in the case of larger molecules, from surface 
friction. (c) In both examples (a) and (b), a top-view of the membrane bilayer 
qualitatively resembles a 2D array of immobilized obstacles (Q), 
through which the mobile particles (M) diffuse.}
\end{center}
\end{figure*}

The above mechanisms share in common that they do not require the membrane to be 
coupled to its environment in any specific way. However, an essential difference 
between model membranes and real cells may well be the fact that the latter are 
coupled to their surroundings. For example, the cytoskeleton of a cell may 
induce a tension on the membrane affecting its domain 
structure~\cite{uline:2012}. Another possible effect -- the one that we focus on 
in this work -- is that the coupling of the membrane to its environment causes 
some of its components to become immobilized. A possible 
mechanism leading to particle immobilization 
in a real cell may be an actin network underlying 
the membrane. The actin network is attached to the membrane via anchoring 
proteins and, since the actin network is disordered, their positions will be 
random. The anchoring proteins are essentially immobile compared to the freely 
diffusing membrane components, and thus act as obstacles for the mobile 
components~[\fig{fig:qd}(a)]. In addition to the anchoring proteins themselves, 
also other molecules can attach to the actin strands, which thereby also become 
obstacles. In fact, the distance between obstacles anchored to the actin strands 
can become very small $d \sim 2-9$~nm~\cite{citeulike:6859704}.

The immobilization of molecules can also arise {\it in vitro} in supported 
membranes\cite{citeulike:6609138, citeulike:6528926, citeulike:6116695, 
citeulike:5806549, citeulike:6127454, citeulike:6598182, citeulike:4071068}. 
Fluorescence measurements have revealed that surface roughness profoundly limits 
lipid diffusion, even leading to complete immobilization. For example, the 
immobilized lipid fraction is around 15\% on alumina substrates, and 5\% on 
silica \cite{citeulike:6116695}. Furthermore, due to surface friction, embedded 
proteins may also become immobilized~\cite{citeulike:6127454} [\fig{fig:qd}(b)]. 
Hence, both in biological membranes as well as in artificially supported lipid 
bilayers, lateral diffusion may be envisioned as taking place inside a random 
network of static obstacles [\fig{fig:qd}(c)]. In physical terms, the presence 
of a random network of static obstacles constitutes a form of quenched 
disorder.

\subsection{Quenched disorder in membranes: \\ Summary of known results}


The above examples show that the presence of quenched disorder in membranes is 
inevitable in many practical situations. This naturally raises the question to 
what extent this can account for the properties of a membrane. Regarding single 
particle diffusion, it is known that lateral diffusion constants are drastically 
reduced in the presence of obstacles \cite{citeulike:918597, citeulike:6505316}. 
The diffusion constant of band 3 in mouse erythrocytes without cytoskeleton, 
i.e.~in the {\it absence} of quenched disorder, is 50 times larger compared to 
that in healthy cells \cite{citeulike:6505316}. Additionally, the diffusion 
becomes anomalous (non-Brownian) on intermediate time-scales. These results can 
be explained by assuming that the quenched obstacles divide the membrane into 
compartments~\cite{citeulike:6859704, citeulike:918597, citeulike:9836137, 
citeulike:5393528, citeulike:6505558, citeulike:6195415, citeulike:6116283, 
citeulike:4304224, citeulike:139732}. Within a single compartment, the diffusion 
is fast and Brownian, while diffusion between compartments is governed by a much 
slower \ahum{hopping} dynamics \cite{citeulike:918597}.


In addition to {\it single} particle diffusion, quenched obstacles also affect 
the {\it collective} behavior of the mobile constituents. By collective behavior 
we mean in this context the size and shape of the liquid-ordered and 
liquid-disordered domains that can form in the membrane. Precisely this question 
was addressed in a series of recent Monte Carlo (MC) 
simulations~\cite{citeulike:7115548, citeulike:6599228, citeulike:8864903, 
Machta20111668, Ehrig201180}. A remarkable finding is that, provided (1) the 
quenched obstacles are randomly distributed, and (2) show a preferred attraction 
to one of the mobile components, macroscopic domain formation is entirely 
prevented~\cite{citeulike:8864903}. This behavior directly connects to the 
fundamentals of fluid phase behavior in quenched porous media, for which 
numerous theoretical results are available \cite{madden.glandt:1988, 
citeulike:4892193, citeulike:10004339}. In particular, when conditions (1) and 
(2) above are met, the obstacles induce a change in the universality class, from 
Ising toward random-field Ising \cite{gennes:1984}. As is well known, the latter 
does not support macroscopic domain formation in two dimensions 
\cite{imry.ma:1975, citeulike:2841682}. Hence, based on the fundamentals of 
statistical physics, particle immobilization provides a robust mechanism to 
account for a heterogeneous membrane domain structure.

\subsection{Aim of paper}

In this paper, we put the recent MC findings \cite{citeulike:8864903, 
Machta20111668, Ehrig201180} pertaining to particle immobilization and its 
effect on membrane domain formation to a stringent test. It should be noted that 
these previous studies are based on rather simple models: all exclude membrane 
height fluctuations, most are defined on a lattice, lipid tail degrees of 
freedom are either ignored or only very crudely modeled, while solvent and 
cholesterol are absent. In addition, since all deployed the MC method, the 
dynamics could only be approximated or was omitted altogether. Hence, it is 
important to verify if the observed trends in the MC studies also survive in 
more realistic membrane models. The aim of this paper is to provide this 
verification. We present a high-resolution Molecular Dynamics (MD) study of a 
phase-separating DPPC/DLiPC/cholesterol lipid bilayer in the presence of an 
immobilized component, which captures important details left out in earlier 
works. 

The outline of our paper is as follows: In Section~\ref{sec:MD}, we present our 
MD simulation results, and show that the presence of quenched obstacles in a 
membrane indeed results in a heterogeneous structure of small domains. In 
Section~\ref{sec:MC}, we then compare our MD results to the predictions of a 
simple MC model (the 2D Ising model), and demonstrate that the latter is already 
well-suited to describe domain structures. We end with a summary and conclusion 
in Section~\ref{sec:summary}.

\section{Molecular Dynamics simulations \label{sec:MD}}

\subsection{MD simulation setup}

For the MD simulations presented here, a setup from a previous 
work\cite{citeulike:3483967} was used, adapted to our needs by changing the size 
and shape of the membrane patch and adding quenched disorder to the system. In 
the setup without disorder, which is simulated first for comparison, the 
membrane consists of $1532$ cholesterol molecules and two species of 
phospholipids, namely $1408$ molecules of dipalmitoyl-phosphatidylcholine with 
fully saturated carbon tails (DPPC), and $2184$ molecules of the polyunsaturated 
phospholipid dilinoleyl-phosphatidylcholine (DLiPC). This composition lies deep 
within the coexistence region of the phase diagram~\cite{VeatchPhasediagram, 
citeulike:5629089, citeulike:9049927}. Hence, our lipids have a strong tendency 
to demix.

\begin{table}
\begin{center}
\begin{tabular}{c|l|l}
Run & Type of obstacles & $t_{\rm max}$ \\
\hline
I   & None & $25$~$\mu$s \\
IIa & Same on both sides, mobile in $z$-direction & $18$~$\mu$s \\
IIb & Same on both sides, mobile in $z$-direction & $22$~$\mu$s \\
III & On one side only, pinned in $z$-direction   & $22$~$\mu$s
\end{tabular}

\caption{\label{tab:sims} Overview of the four MD simulation runs performed in 
this work, listing the immobilized obstacle type, and the total simulation 
time~$t_{\rm max}$. Runs IIa and IIb use the same obstacle configuration, but 
different random numbers for the initialization of the mobile lipids were used. 
Note that diffusion rates in the \texttt{MARTINI} force field are $\sim 4$ times 
larger compared to atomistic simulations and 
experiments~\cite{doi:10.1021/jp071097f}. Values for time given in this work 
refer to the simulation time coordinate, not to the rescaled (i.e.~larger) 
\ahum{physical time}.}

\end{center}
\end{table}

In the initial state, the lipids form a flat bilayer patch in the $(x,y)$ plane, 
using a periodic simulation box of size $45 \times 30 \times 11$~nm with the 
same number of DPPC and DLiPC lipids in both layers. Within each layer, the 
phospholipids are initially randomly mixed. The molecules are modeled with the 
\texttt{MARTINI} force field\cite{doi:10.1021/jp071097f}, which provides a 
near-atomic resolution of $3$-$4$ non-hydrogen atoms per simulation bead. In 
order to capture the effect of the solvent, $74822$ solvent beads were included. 
The simulation has been performed with 
\texttt{GROMACS}~\cite{JCC:JCC20291,doi:10.1021/ct700301q} Version~$4$ at 
temperature $295$~K and a pressure of $1$~bar (for further details we refer the 
reader to the original setup~\cite{citeulike:3483967}). In total, four different 
simulation runs have been performed, summarized in \tab{tab:sims}, and to be 
discussed further in the following sections.

\subsection{The free membrane \label{sec:pure}}

We first consider the \ahum{free} membrane (run I in \tab{tab:sims}), by which 
we mean a membrane without any quenched disorder. In the absence of quenched 
disorder, model ternary membranes of phospholipids and cholesterol readily phase 
separate into macroscopic ($> \mu$m) domains~\cite{Veatch20033074}. The domains 
observed in experiments are typically round, as this shape minimizes the length 
of the line interface. The key difference with computer simulations using 
periodic boundary conditions is that the equilibrium domain shape is {\it not} 
necessarily round. This is a crucial point in our analysis and to avoid possible 
confusion with related studies~\cite{Ehrig201180, citeulike:9234255} we explain 
it explicitly. In a 2D system with periodic boundary conditions, the two 
candidate domain shapes are (1) a circle, or (2) a stripe that spans the system 
along the shorter edge of the simulation box~\cite{citeulike:7134501, 
citeulike:7237424}. The shape that prevails is the one with the shortest line 
interface, which depends on the area fraction of the two coexisting phases and 
the aspect ratio of the simulation box. For the simulations presented here, the 
hallmark of macroscopic domain formation in a {\it finite} simulation box with 
periodic boundaries is the appearance of the latter stripe geometry. 
Consequently, in our MD simulations, the appearance of such a stripe domain will 
be regarded as evidence for macroscopic domain formation.

\begin{figure}
\begin{center}
\includegraphics[width=0.95\columnwidth]{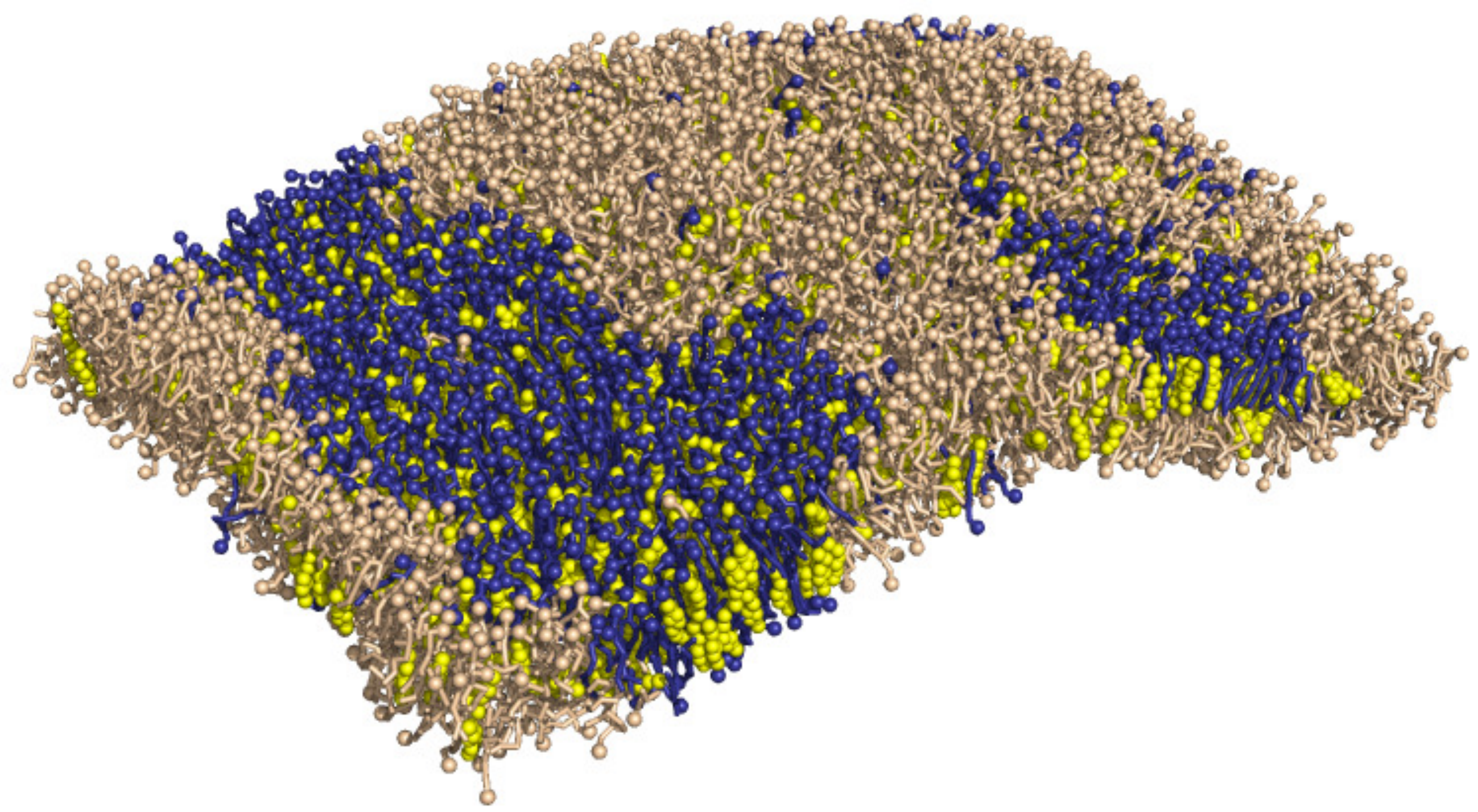} 
\end{center}
\caption{ \label{fig:pure} \small
  Membrane without quenched 
  disorder (run~I)
  at $t=5$~$\mu$s, showing the upper 
  leaflet. Due to inter-leaflet
  coupling, the domain structure on the lower leaflet is
  almost identical.
  From the initially mixed state, two domains of DPPC 
  (dark blue) and cholesterol (yellow) 
  have formed in a background of DLiPC (bright brown).
  The DPPC domain in the foreground has
  assumed the stripe shape, which is the simulation analogue of large round 
domains seen in experiments.}
\end{figure}


We started the MD simulation of the free membrane with a mixed configuration. 
During the first few $\mu$s of simulating the membrane, we observe the 
characteristics of binary demixing, namely the formation, coarsening, and 
merging of domains. These domains reflect the liquid-ordered phase (rich in DPPC 
and cholesterol) and the liquid-disordered phase (rich in DLiPC). Due to 
inter-leaflet coupling\cite{citeulike:3483967}, domains in both leaflets arrange 
in similar shapes and at the same locations. After $5$~$\mu$s, one of the 
domains has assumed the stripe geometry, which remains stable for the rest of 
the simulation [\fig{fig:pure}]. Following the explanation above, this indicates 
macroscopic demixing into large DPPC-rich and DLiPC-rich domains, such as seen 
experimentally in free-standing giant unilamellar 
vesicles~\cite{Bagatolli2000290, Veatch20033074}. Ultimately, the small DPPC 
domain on the right of \fig{fig:pure} will merge with the stripe, but this 
process is very slow and beyond our computational resources.
	
\subsection{Membranes with quenched disorder}

\begin{figure}
\begin{center}
\includegraphics[width=0.7\columnwidth]{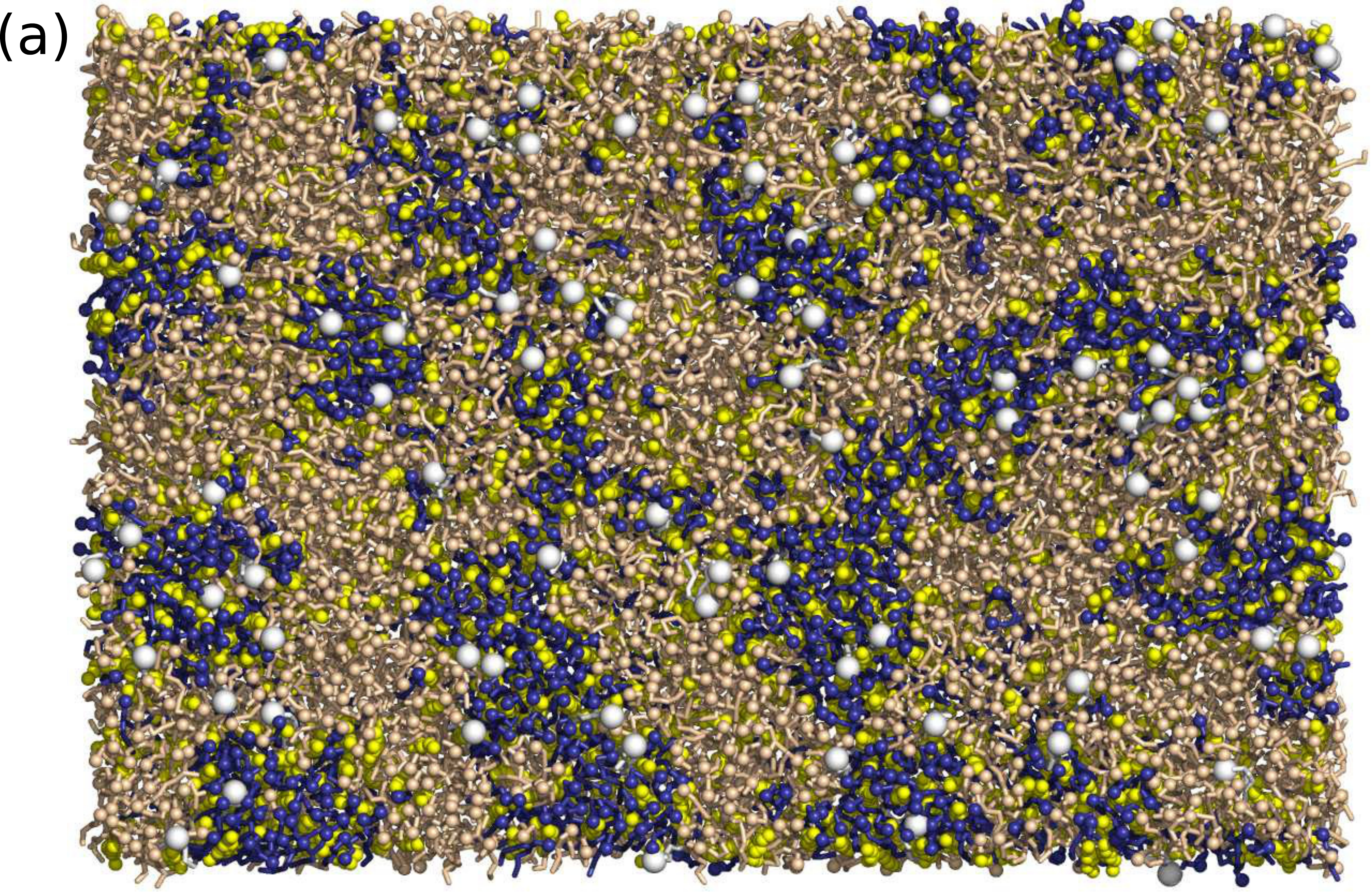}\\
\includegraphics[width=0.7\columnwidth]{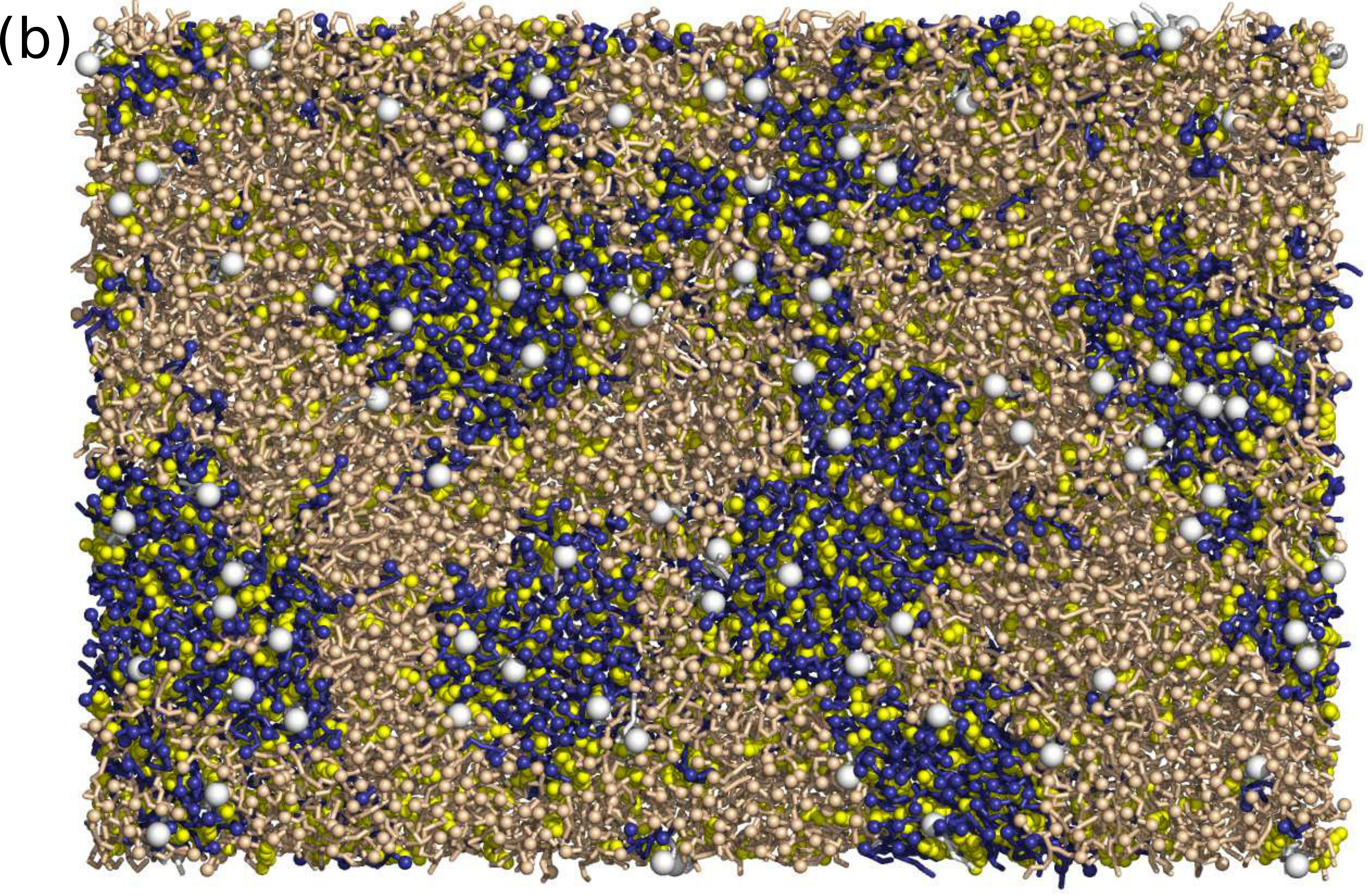}\\
\includegraphics[width=0.7\columnwidth]{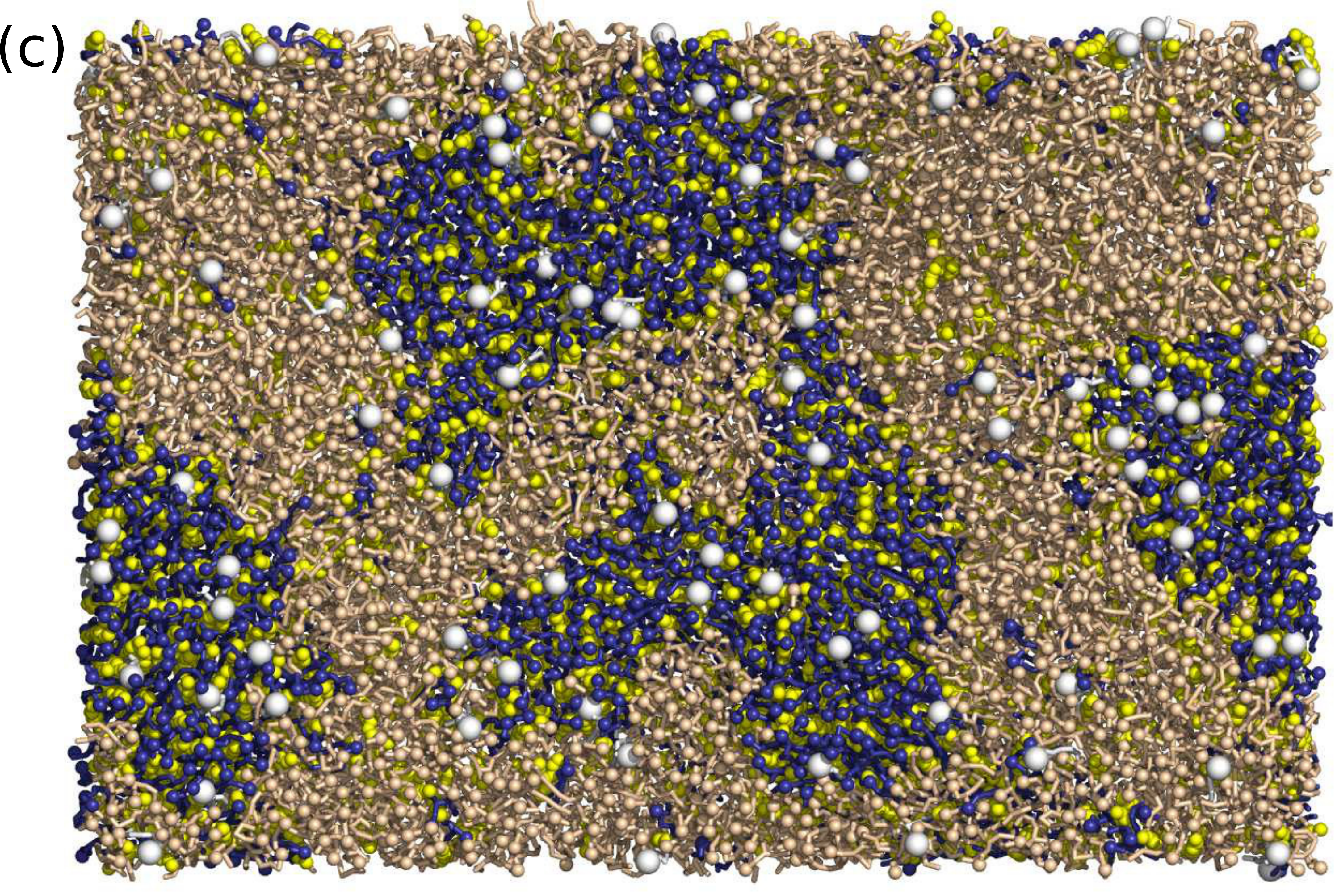}
\end{center}
\caption{\label{fig:domainformation} \ahum{Early time} domain structures after 
$t=0.5$~$\mu$s (a), $t=2$~$\mu$s (b), and $t=5$~$\mu$s (c) as observed in 
simulation run IIa (cf.~\tab{tab:sims}). The obstacles are shown as white; 
the remaining color-coding is the same as in \fig{fig:pure}.}
\end{figure}

\begin{figure*}
\begin{center}
\includegraphics[width=0.47\textwidth]{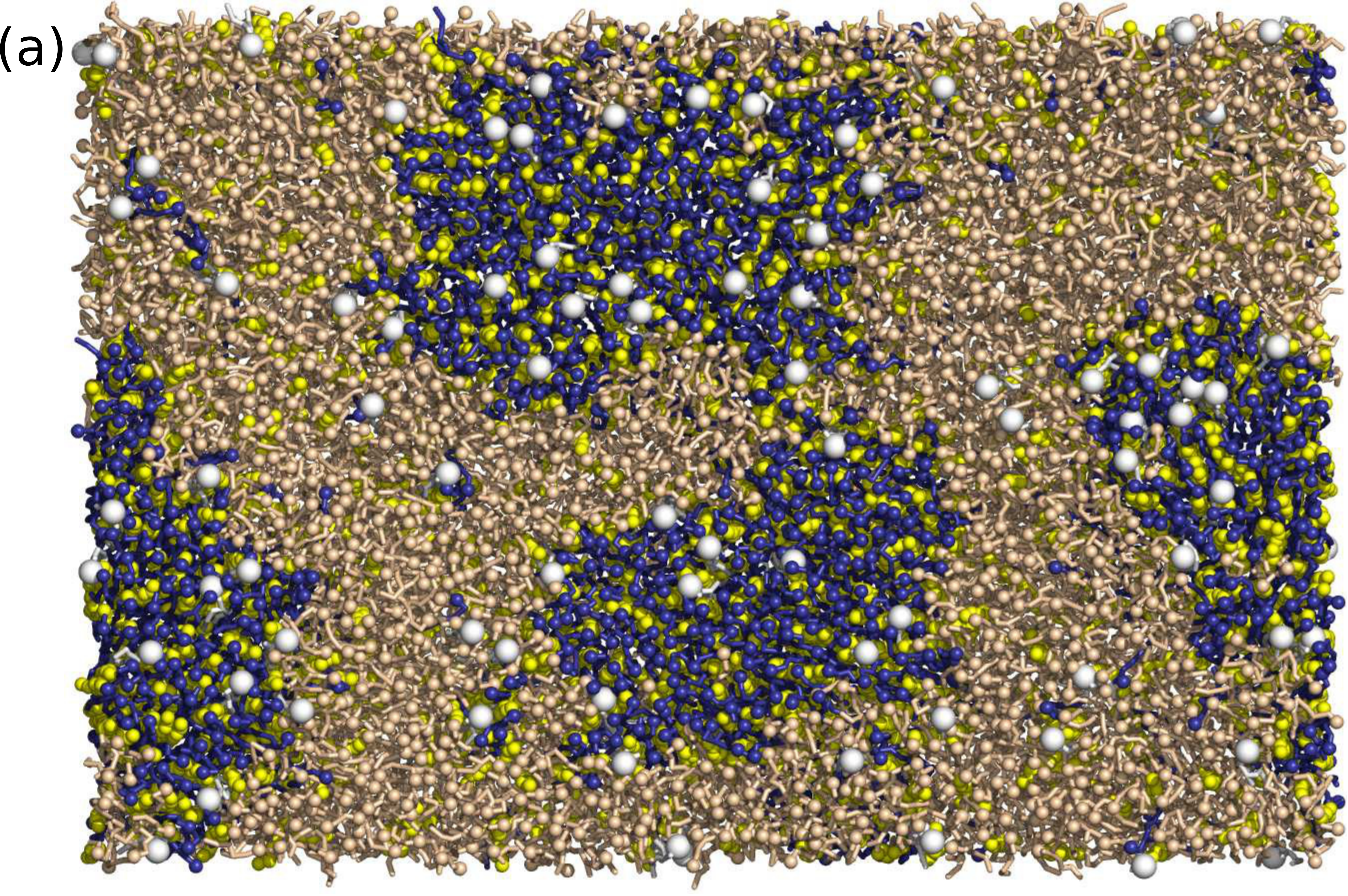}
\hspace{0.02\textwidth}
\includegraphics[width=0.47\textwidth]{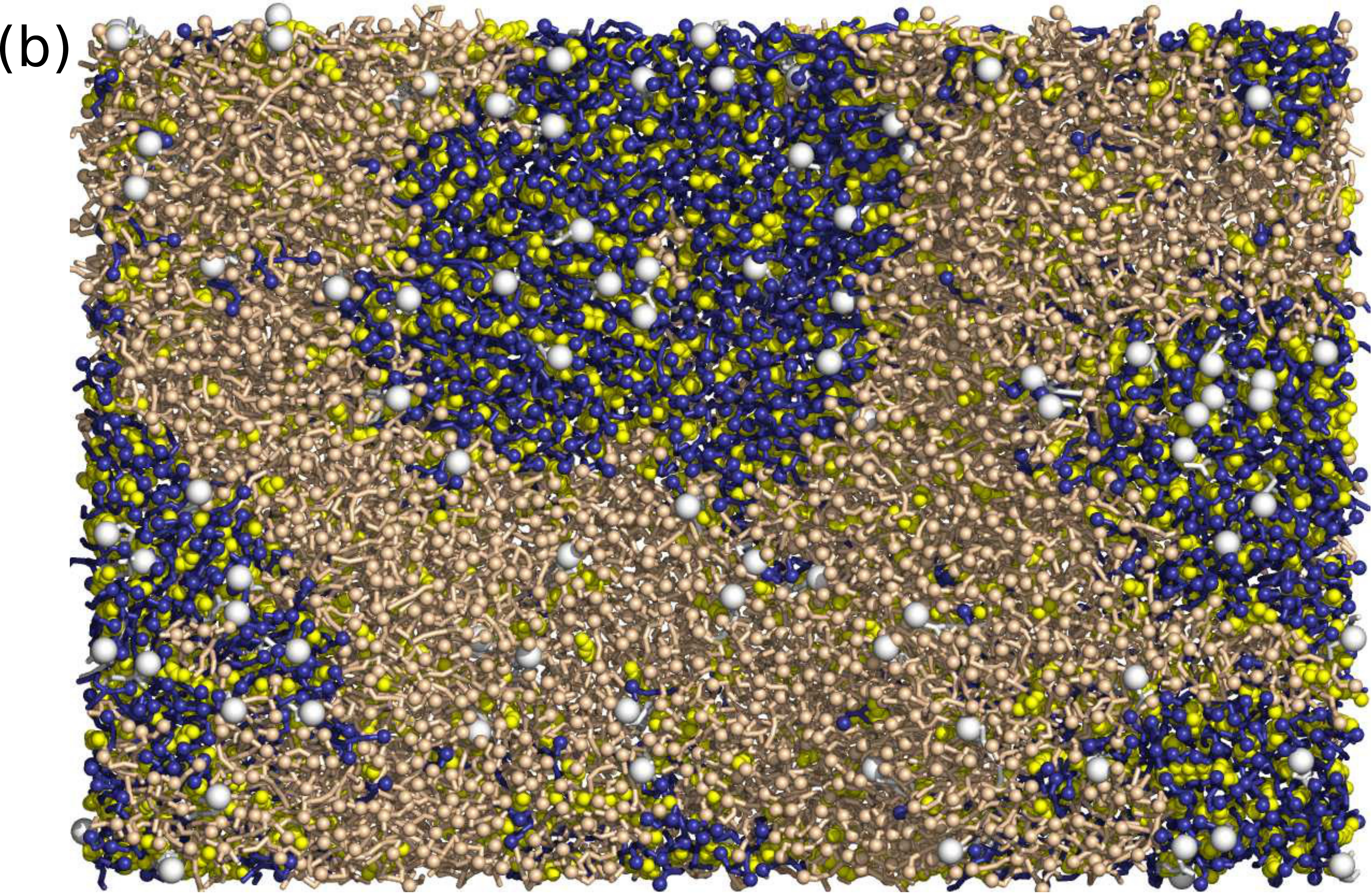}
\end{center}
\caption{\label{fig:obstacles} \ahum{Late time} domain structures of
runs IIa (a) and IIb (b) at $t=10$~$\mu$s. FPPC molecules are shown as white and 
are at the same locations in both systems (the slightly different positions of 
the white beads between the snapshots result from only the phosphate atom of the 
FPPC being quenched). Otherwise, the color-coding is the same as in 
\fig{fig:pure}. In both simulations, the stripe domain characterizing 
macroscopic domain formation is absent. Instead, structures consisting of small 
domains are seen.}
\end{figure*}


To demonstrate how quenched disorder affects structure formation in membranes, 
obstacles have been added to the free membrane of Section~\ref{sec:pure}. 
Contrary to previous studies of membranes with quenched 
disorder\cite{citeulike:7115548, citeulike:6599228, citeulike:8864903, 
Machta20111668, Ehrig201180} that only took into account a flat monolayer our 
model takes into account the full bilayer and allows for membrane undulations. 
Hence, there are two additional degrees of freedom we can investigate. First of 
all, the obstacles may be allowed to follow the membrane undulations in the 
$z$-direction, or be fixed in this direction. Secondly, the obstacles may be 
placed at the same lateral locations in both leaflets, which may be conceived to 
describe a trans-membrane protein, or they could be chosen to reside on only one 
leaflet.

\subsubsection{Obstacles that follow membrane fluctuations  \label{sec:obstacles}}


We first consider the case whereby the obstacles have been placed at the same 
locations in both leaflets and are allowed to move freely in the $z$-direction. 
In this scenario, the obstacles correspond to trans-membrane proteins that are 
free to move in the $z$-direction themselves, or so long that the membrane can 
fluctuate around them. To model the obstacles, we chose to take DPPC molecules 
that are not allowed to diffuse laterally. Such obstacles, which we refer to as 
FPPC molecules in what follows, naturally attract the DPPC-rich phase. The 
motivation to use DPPC obstacles stems from experimental considerations: 
cytoskeleton anchor proteins typically favor the liquid-ordered 
phase~\cite{citeulike:10759324, Ehrig201180}. In principle, we could have chosen 
DLiPC-affine obstacles or a mix of DPPC-affine and DLiPC-affine obstacles. 
Similar results as those presented here would be expected in these cases also.


To generate an initial state of a membrane with obstacles, we take the initial 
state from Section~\ref{sec:pure} and chose $79$ random lateral locations in the 
membrane. Then, in each leaflet the $79$ nearest phosphate atoms belonging to a 
DPPC molecule are identified and moved to the target positions by a series of 
small translations and subsequent energy minimizations (so that the DPPC 
molecules remain intact and that there is no overlap between lipids). Finally, 
the so-moved DPPC molecules are labeled FPPC and the $(x,y)$-coordinates of 
their phosphate atoms are bound to their target locations via a hard harmonic 
potential. During the course of the simulation, the lateral positions to which 
the phosphates of the FPPC are bound scale with the volume changes of the 
simulation box. Otherwise they are static, effectively making the FPPC quenched.


Two simulation runs with the same FPPC obstacle configuration have been 
performed (IIa and IIb of \tab{tab:sims}), both starting from a state of 
\ahum{mixed} mobile lipids. In both simulations, a formation and coarsening of 
small DPPC-rich domains is seen during the first few microseconds (as shown in 
\fig{fig:domainformation} for run~IIa). After approximately $5$~$\mu$s, however, 
domain growth and merging of domains into larger ones is arrested. Beyond this 
time, the domains exhibit shape fluctuations, but do not significantly change in 
size or location for the rest of the simulation. Most importantly, as shown in 
\fig{fig:obstacles}, neither of the simulation runs reveals the stripe geometry 
seen in \fig{fig:pure} that would indicate domains larger than the size of the 
simulation box. We conclude that the obstacles indeed prevent macroscopic 
demixing, and replace it with the formation of small domains that appear stable 
over times at least in the order of $\sim 10$~$\mu$s.


In recent MC studies, it was shown that quenched obstacles induce regions in the 
membrane where either a liquid-ordered or a liquid-disordered domain is 
preferred\cite{citeulike:8864903, Machta20111668}. At zero temperature this 
would essentially fix the domain structure into the groundstate configuration, 
while at finite temperature thermal fluctuations still allow some freedom as to 
where the domains form. This explains why, despite identical FPPC locations, the 
domain structures in \fig{fig:obstacles}(a) and (b) are not the same. We return 
to this point in detail in Section~\ref{sec:MC}.

\subsubsection{Obstacles on a single leaflet \label{sec:singlesided}}

\begin{figure}
\begin{center}
\includegraphics[width=0.9\columnwidth]{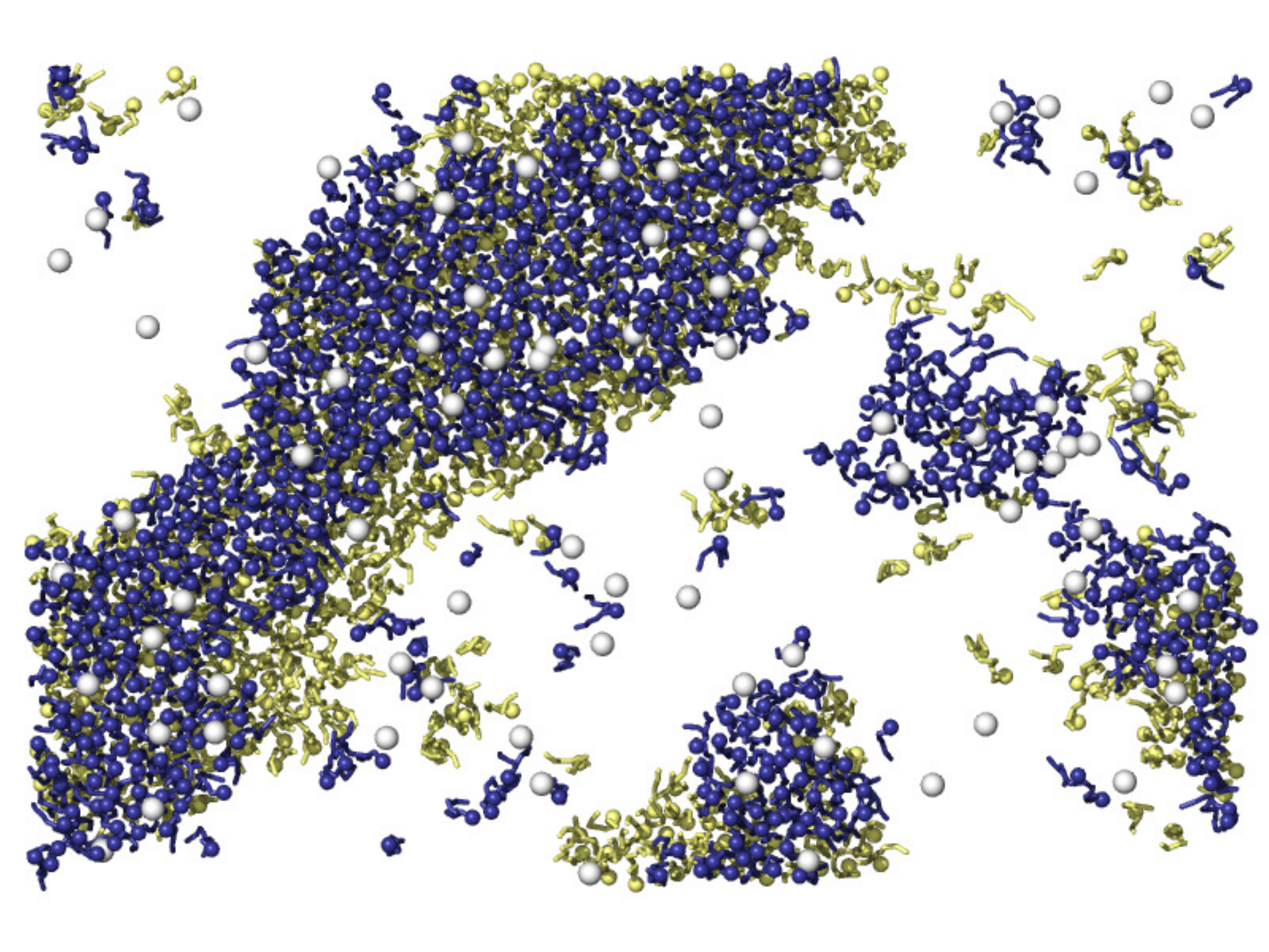}
\end{center}

\caption{\label{fig:single} Top-view of a membrane with FPPC obstacles in only 
one bilayer leaflet (run~III) at $t=10$~$\mu$s. For clarity, this figure does 
not show the DLiPC or cholesterol molecules. DPPC belonging to the leaflet with 
the FPPC are drawn in blue, those of the \ahum{free} leaflet in yellow. The 
domains form \ahum{in registry} in both leaflets, but they are larger compared 
to the case where FPPC resides in both leaflets. The structure is dominated by a 
single large domain in both leaflets, but it is distinctively different from the 
stripe geometry that would indicate macroscopic domains.}

\end{figure}

We now consider the case where the FPPC obstacles reside in only one of the 
bilayer leaflets, and do {\it not} follow the membrane height fluctuations 
(run~III). Such obstacles thus locally \ahum{pin} the membrane height, which 
could mimic a cytoskeleton anchoring site or, in the case of an adhered 
membrane, a (stiff) ligand-receptor bond~\cite{citeulike:10571682}. This 
scenario is physically interesting because it probes to what extent 
inter-leaflet coupling transmits the influence of quenched disorder from one 
leaflet to the other. To address this scenario in our MD simulations, we perform 
the same setup steps as in Section~\ref{sec:obstacles} with the same $79$ target 
locations for the obstacles, but we place the obstacles on only one of the 
leaflets and also bind the $z$-coordinate of the FPPC phosphate atoms to the 
hard harmonic potential.

Our simulations indicate that inter-leaflet coupling is indeed an important 
mechanism in domain formation, which is consistent with previous 
studies~\cite{putzel:2011, schick:2011}. During the first few microseconds, the 
formation, coarsening, and merging of domains is observed in both leaflets. The 
domains form \ahum{in registry} in both leaflets, occurring at the same 
(lateral) positions. After $10$~$\mu$s, most of the DPPC molecules are contained 
in a single domain as shown in \fig{fig:single}. We thus obtain larger domains 
compared to the case of FPPC residing in both leaflets. At the same time, we do 
not obtain the stripe geometry, suggesting that the domains are not macroscopic. 
\Fig{fig:single} thus presents a \ahum{hybrid} structure, between the stripe 
domain of \fig{fig:pure} and the small domains of \fig{fig:obstacles}. This 
result shows that the effect of quenched disorder is transmitted from the 
leaflet with FPPC to the one without, but also indicates that the tendency of 
the leaflet without FPPC to form large domains \ahum{back-transmits}. The net 
result is a structure of finite-sized domains (i.e.~not macroscopic) but of a 
typical size that exceeds the case where FPPC resides in both leaflets.

\subsection{Interface properties of domains}

\begin{figure}
\begin{center}
\includegraphics[width=0.9\columnwidth]{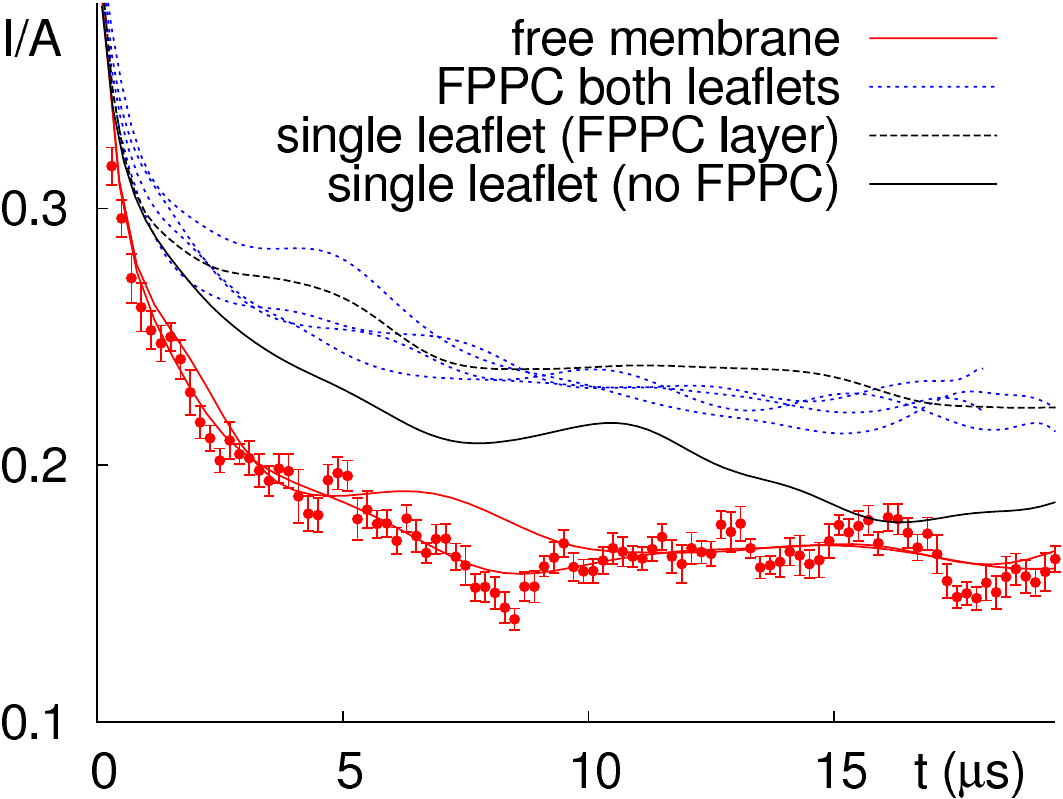}
\end{center}
\caption{\label{fig:domainsizes} Ratio of the interface length $I$ and domain 
area $A$ of the largest domain for the free membrane (run~I), the membrane with 
FPPC in both leaflets (runs~IIa and IIb), and for a membrane with FPPC in only 
one leaflet (run~III). For each run, results are shown for each leaflet 
separately (hence 8 curves in total). Already at early times, $I/A$ for the free 
membrane differs from the membranes with quenched disorder, and saturates at a 
significantly lower value. All curves shown are interpolations of the raw 
simulation data. For clarity, the raw data are only shown explicitly for one 
curve. The scatter in the raw data of the other curves is 
comparable.}
\end{figure}


We now consider a more quantitative measure to distinguish membranes with 
quenched disorder from those without. To this end, we consider the ratio $I/A$ 
of the domain interface length $I$ to domain area $A$. In the free membrane, the 
formation of macroscopic domains is driven by the desire of the system to 
minimize the interface length, yielding a low value of $I/A$. In the membrane 
with quenched disorder, where stable small domains are expected, $I/A$ will be 
larger. To measure $I/A$, a $153 \times 102$ grid of rectangles is superimposed 
on the membrane. The positions of the phosphate atoms of the DPPC molecules are 
then projected onto the grid. If a grid cell contains the phosphate atom of a 
DPPC molecule, then that cell is considered to be part of a DPPC-rich domain. 
Cells with less than four neighbors belonging to a DPPC-rich domain are defined 
as interface.

In \fig{fig:domainsizes}, the ratio of the number of interface squares $I$ to 
the number of total squares $A$ for the largest DPPC-rich domains are shown for 
both leaflets and for all our simulations. In all cases, the curves first 
show a rapid drop resulting from the initial formation of domains, followed by a 
saturation as the system equilibrates. The saturation values for the free 
membrane, however, are distinctly below those of the membranes with quenched 
disorder. This is consistent with macroscopic domain formation in the former, 
and stable finite domains in the latter. Note also that, for the membrane with 
FPPC obstacles in only one leaflet, $I/A$ is largest in the FPPC containing 
leaflet.

\subsection{Choice of order parameter}

In our analysis, domains were identified by their phospholipid content. An 
alternative choice is to identify domains via the tail ordering of the 
phospholipids, i.e.~the identification of liquid-ordered and liquid-disordered 
domains~\cite{citeulike:9234255}. To this end, we have also quantified the 
elongation of the phospholipid tails via a nematic order parameter. DPPC and 
DLiPC lipids typically have high and low nematic order parameters, respectively. 
Simulation snapshots in which the phospholipids are color-coded according to 
their nematic order parameter reveal the same domain structures as those 
presented here. The identification of domains according to lipid content or tail 
conformation is therefore equivalent, and will not be explicitly presented here.

\section{Comparison to the 2D Ising model \label{sec:MC}}


In Section~\ref{sec:MD}, we have shown using a detailed membrane model that 
quenched obstacles prevent the formation of large (macroscopic) domains. This 
confirms the validity of the trends observed in earlier MC studies on simple 
membrane models~\cite{citeulike:6599228, citeulike:8864903, Machta20111668, 
Ehrig201180}. One issue that remains open is whether the apparent elimination of 
macroscopic domains really is the correct equilibrium behavior, or merely caused 
by the obstacles slowing down the dynamics of domain formation such that the 
relevant timescales become inaccessible in the simulation. Because of their high 
computational demand\footnote{Each of the MD runs of \tab{tab:sims} required 
several months of simulation with 32 processors in parallel.} our MD simulations 
alone cannot resolve this issue. However, in MC simulations of simple models, 
where the proper equilibrium behavior (of these simple models) can be 
accessed\footnote{In the case of MC simulations of the 2D Ising model, the 
equivalent of one MD run takes only a few seconds on a single processor.}, it is 
predicted that macroscopic domain formation is not merely delayed, but in fact 
eliminated (the theoretical motivation is the Imry-Ma 
argument~\cite{imry.ma:1975, citeulike:8864903}). In this section, we will show 
that these MC models make further predictions that can be tested against our MD 
results. Provided one uses the same positions for the obstacles, the resulting 
domain structure is remarkably insensitive to the model details. In fact, the 
minimal model requirement is that the unlike lipid species repel each other, 
which is already contained in the 2D Ising model. In what follows, we will show 
that the domain structures obtained in \fig{fig:obstacles} with MD, can indeed 
be reproduced in MC simulations of the 2D Ising model.

In the 2D Ising model, each lattice site~$i$ of a 2D square lattice is occupied 
by a spin variable $s_i = \pm 1$. In the context of membranes, $s_i=+1 \, (-1)$ 
means that site~$i$ contains a DPPC (DLiPC) lipid, while cholesterol is ignored 
on this level~\cite{Machta20111668}. The energy of an Ising spin configuration 
is given by $E = -J \sum_{<i,j>} s_i s_j$, with coupling constant $J>0$ (the sum 
runs over nearest neighbors). We use the Ising model to mimic the domain 
structures of \fig{fig:obstacles} by simulating a $51 \times 34$ lattice. This 
approximates the aspect ratio of the MD simulation box, as well as the total 
number of phospholipids in a single leaflet. To represent the FPPC obstacles, 
the lattice is superimposed on the initial state of the MD simulations with 
obstacles. Those sites containing an FPPC phosphate atom subsequently have their 
spin value frozen to $s_i=+1$. The remaining sites are randomly initialized with 
values $s_i=\pm 1$, under the constraint that the resulting (DPPC:DLiPC) 
composition matches that of the MD simulation as well as possible. 

We then use MC simulations with Kawasaki dynamics~\cite{newman.barkema:1999} to 
determine equilibrium domain structures. That is, in each MC move, two non-FPPC 
lattice sites are chosen at random. Then, the corresponding spin values are 
exchanged with the Metropolis probability, $P_{\rm acc} = \min[1,\exp(-\Delta E/ 
\kT)]$, with $\Delta E$ the energy difference caused by the swap, $\kb$ the 
Boltzmann constant, and $T$ the temperature. In the absence of quenched 
disorder, the Ising model exhibits macroscopic phase separation when $J / \kT > 
\frac{1}{2} \ln(1+\sqrt{2}) \approx 0.44$, as was shown by 
Onsager~\cite{PhysRev.65.117}. Note that $J / \kT$ is the only free parameter in 
the context of this work.

\begin{figure}
\begin{center}
\includegraphics[width=0.48\columnwidth]{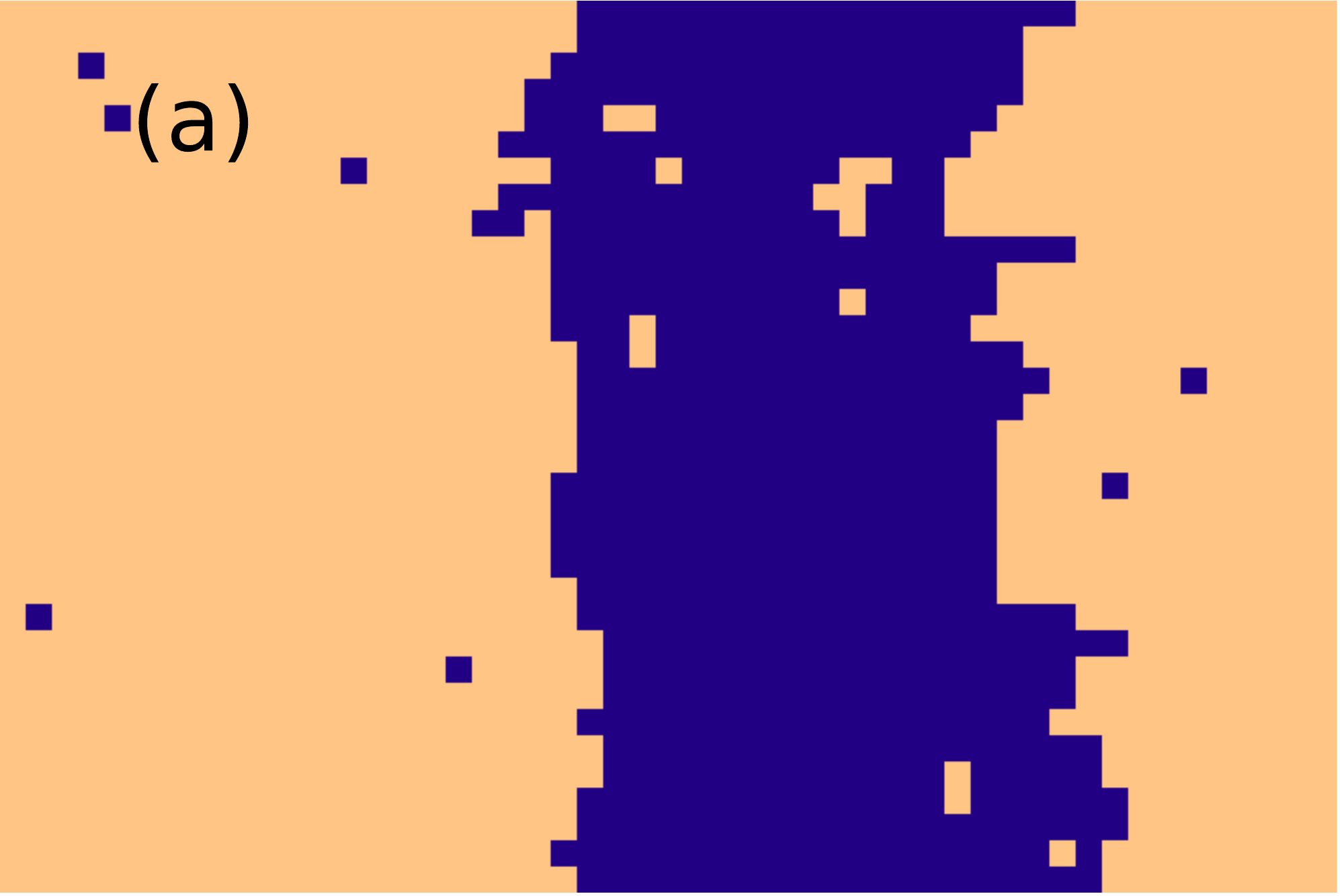}
\includegraphics[width=0.48\columnwidth]{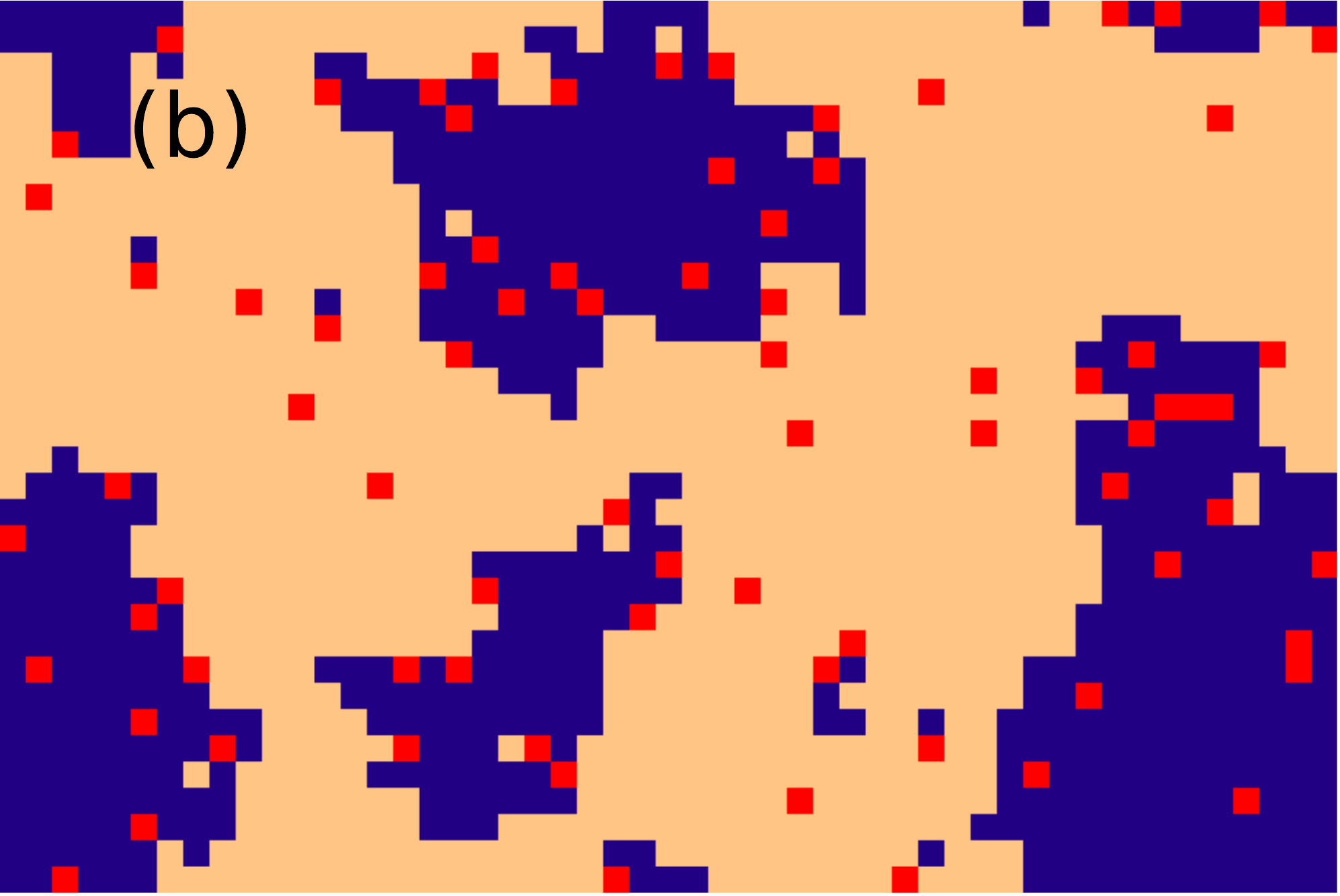} 
\\
\vspace{0.02\columnwidth}
\includegraphics[width=0.48\columnwidth]{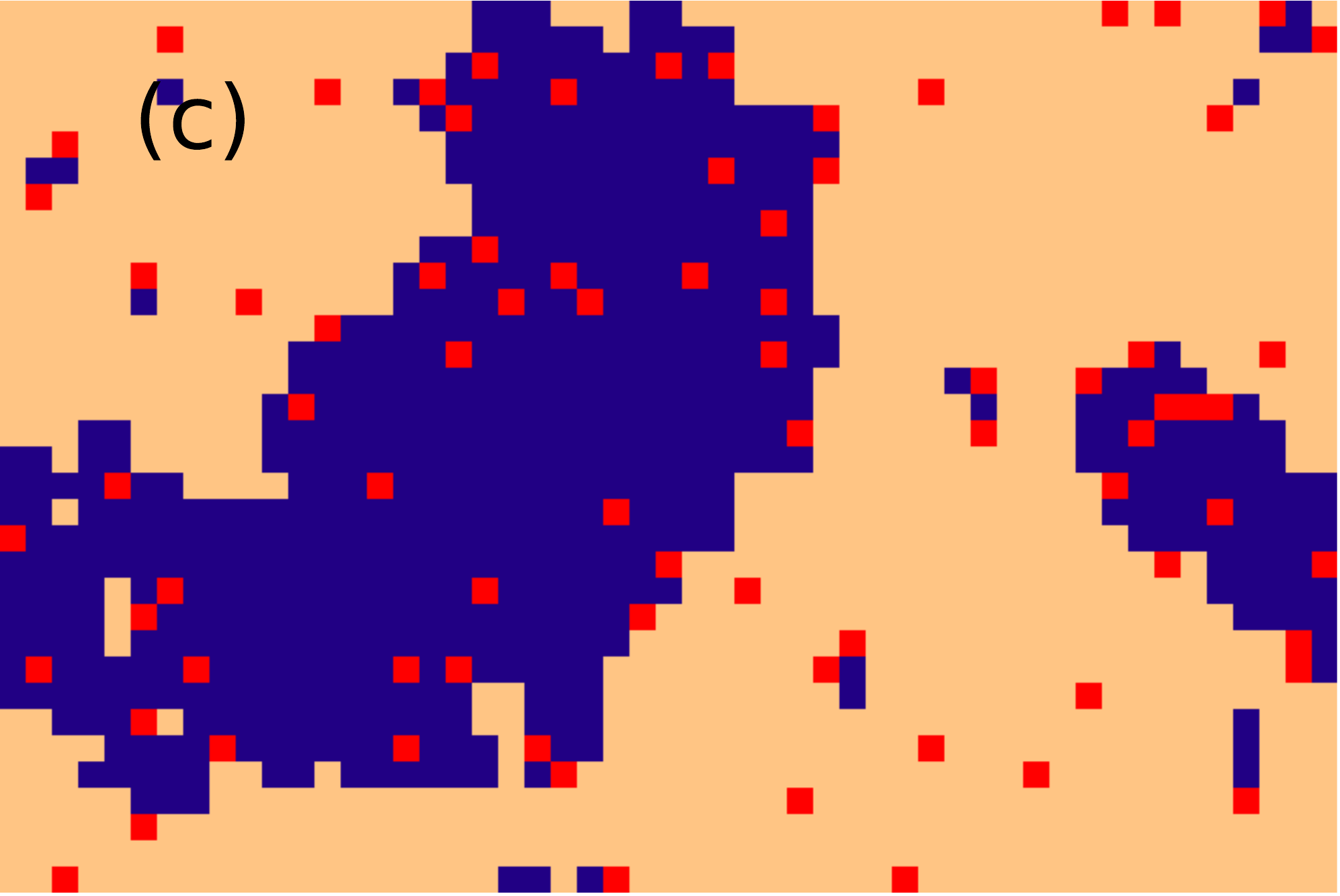}
\includegraphics[width=0.48\columnwidth]{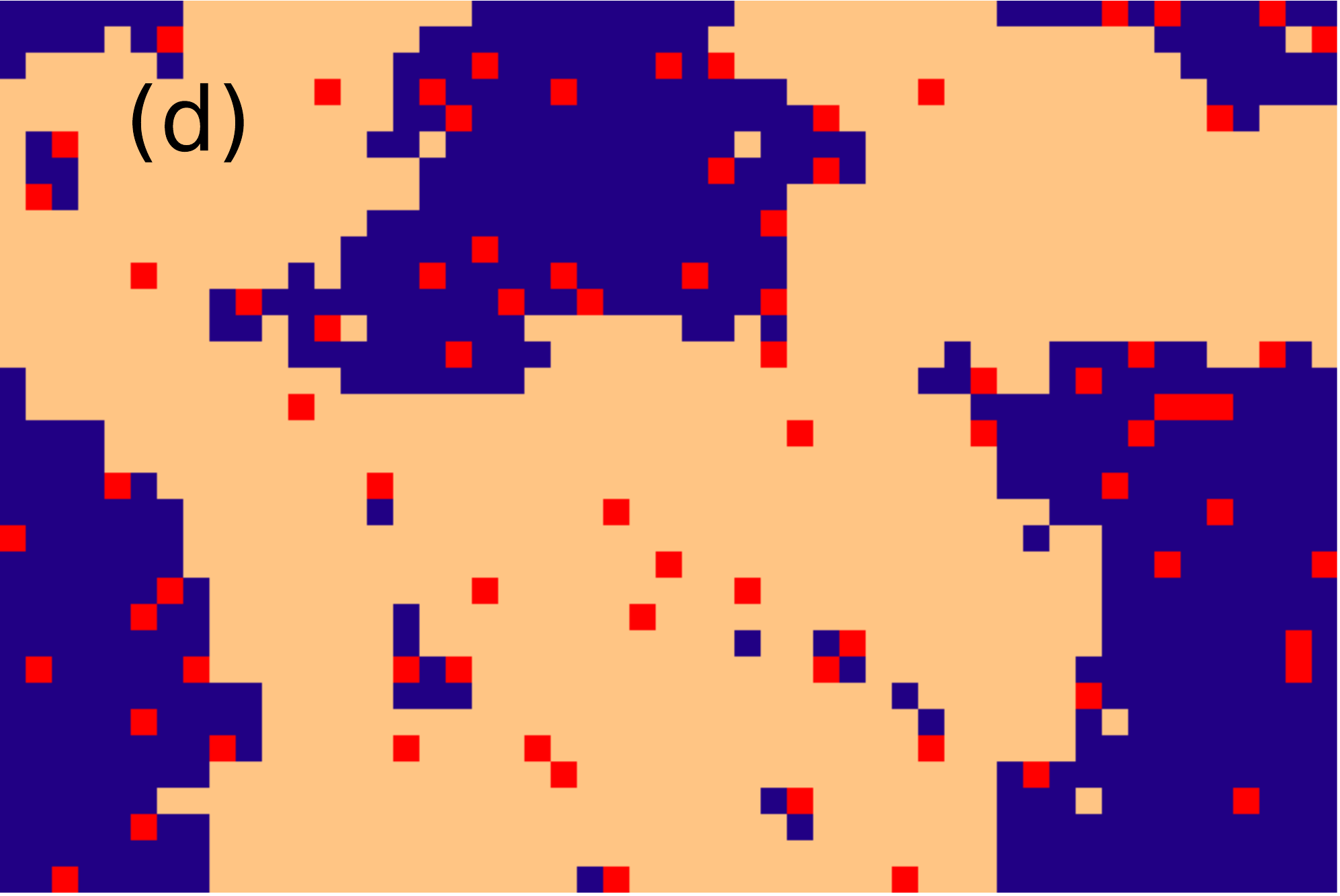}
\end{center}

\caption{\label{fig:IsingStates} Typical domain structures of the \ahum{Ising} 
membrane MC simulation at $J / \kT =0.6$. (a) Domain structure in the {\it 
absence} of quenched obstacles, and with $681$ DPPC (blue) and $1053$ DLiPC 
lipids (orange). The stripe geometry is recovered, in agreement with the 
discussion of Section~\ref{sec:pure}. The snapshots (b,c,d) show typical 
structures in the presence of $78$ FPPC obstacles (red). In each of these 
snapshots, the FPPC configuration is the same, and chosen to reflect the MD 
configuration of \fig{fig:obstacles} (runs IIa and IIb). The differences in the 
snapshots (b,c,d) reflect the thermal fluctuations.}

\end{figure}

At $J/\kT=0.6$, the parameter we have chosen to use for the comparison with the 
MD results, the system therefore assumes the stripe geometry 
[\fig{fig:IsingStates}(a)]. In the presence of quenched disorder, macroscopic 
phase separation is eliminated [\fig{fig:IsingStates}(b,c,d)]. Instead, finite 
domains that {\it qualitatively} resemble those of \fig{fig:obstacles} are seen. 
This similarity is quite remarkable, given that the simulation models used are 
entirely different (only the positions of the obstacles are the same). It shows 
that the generic features of the domain structure are essentially encoded in the 
obstacle configuration, and rather insensitive to the details of the 
interactions. To put this statement on a more solid basis, we now consider a 
{\it quantitative} measure of the correspondence between the domain structures 
seen in the two models.

\subsection{Correlation between the models}

\begin{figure}
\begin{center}
\includegraphics[width=0.9\columnwidth]{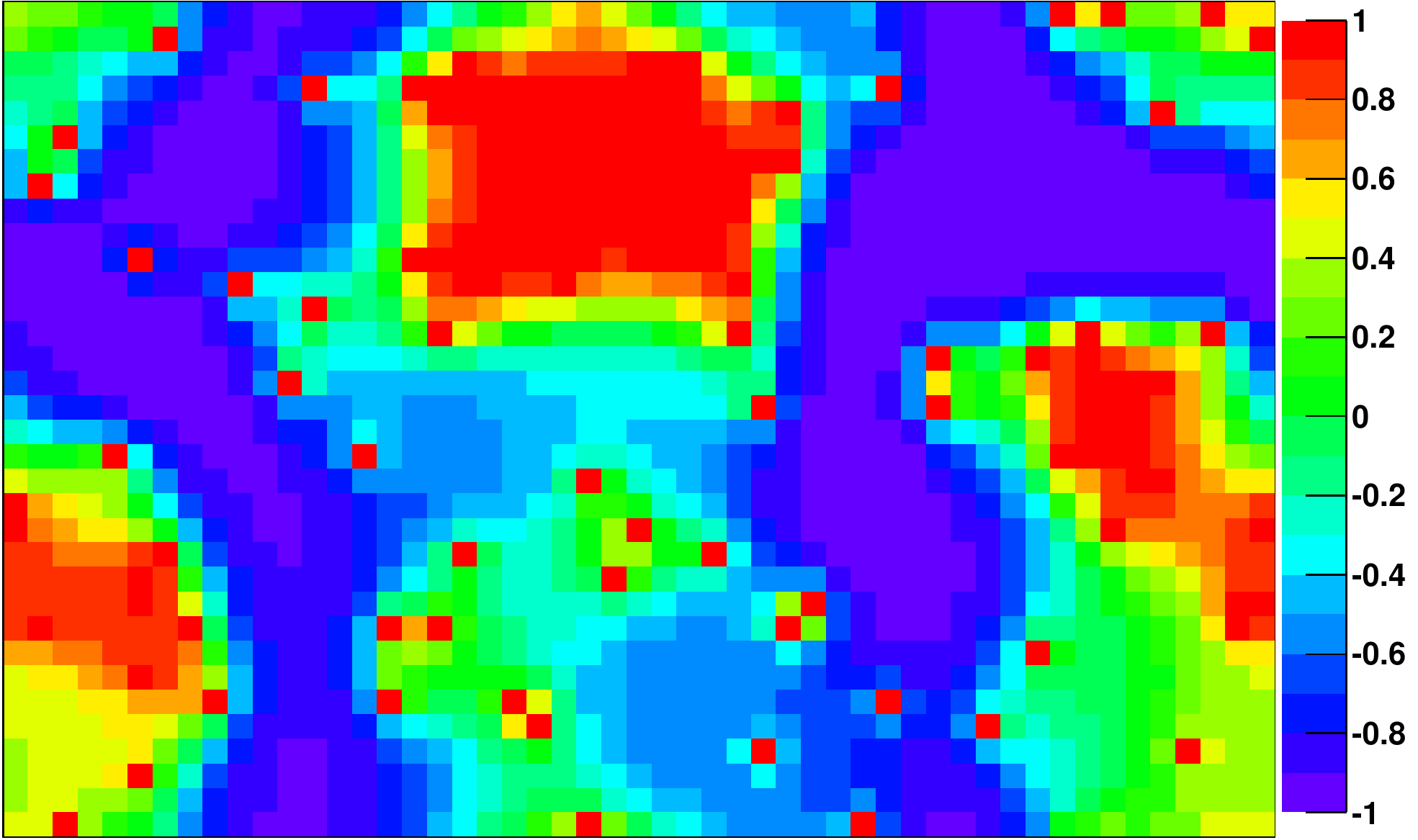}
\end{center}
\caption{\label{fig:Avalues} The {\it ensemble averaged} snapshot (i.e.~the 
$A_i$'s) of the \ahum{Ising} membrane simulation corresponding to the obstacle 
configuration of \fig{fig:IsingStates}(b,c,d) (which, in turn, was derived from 
the MD obstacle configuration of \fig{fig:obstacles}). Preferential locations 
for the appearance of domains are seen, which coincide well with the domain 
structures obtained in the MD simulations of \fig{fig:obstacles}.}
\end{figure}


Because of the Ising model's computational simplicity, many different 
equilibrium states can be generated quickly by means of MC simulation. In fact, 
for the lattice size used here, a representative set of the full equilibrium 
ensemble can be generated. This allows us to create an ensemble averaged 
snapshot by generating a lot of snapshots such as shown in 
\fig{fig:IsingStates}(b,c,d) and evaluating the average spin value $A_i$ at each 
site $i$. As shown in \fig{fig:Avalues}, this average snapshot reveals 
structure, which is due to the fact that the obstacles induce regions that 
locally prefer one of the lipid species~\cite{citeulike:8864903, 
Machta20111668}. With this average structure, we demonstrate the correspondence 
between the MD simulation results and those of the Ising model in two steps: 
First, we define a compatibility measure $C$ that quantifies how much a given 
snapshot of either model resembles the average spin values $A_i$. Then, we 
evaluate the range of compatibilities within the Ising model itself, and 
compatibilities obtained from a structure of domains at random locations 
(i.e.~we want to rule out that an observed correspondence is due to pure 
chance). It will be shown that the compatibilities of our MD results fall into 
the region of the former, and distinctively disagree with the predictions for 
the latter.


We define the compatibility of an Ising model snapshot with the $A_i$ as
\beq \label{def:C}
  C = \frac 1N \sum_i s_i \, A_i \quad, 
\eeq 
where the summation runs over the $N$ lattice sites that do not contain an 
obstacle. This measure quantifies the degree of correlation between a single 
snapshot (characterized by spin values $s_i$) and the ensemble-averaged snapshot 
of \fig{fig:Avalues} (characterized by averages $A_i$). 
To calculate the compatibility of an MD snapshot 
with the averaged Ising structure of \fig{fig:Avalues}, we formally use 
\eq{def:C}, but with proper MD equivalents of the appearing terms. We consider 
each leaflet of the membrane separately, so the sum in \eq{def:C} runs over all 
DPPC and DLiPC molecules in the leaflet, where $N$ is their total number. If 
lipid~$i$ is a DPPC (DLiPC) lipid, then the equivalent spin value $s_i=+1$ 
($s_i= -1$). The corresponding value of $A_i$ is determined by projecting the 
leaflet onto the $(x,y)$ plane, and to then superimpose the averaged Ising 
structure of \fig{fig:Avalues}. For each lipid~$i$, the corresponding $A_i$ is 
taken from the grid cell in the averaged structure that contains the lipid's
center of mass.


\begin{figure}
\begin{center}
\includegraphics[width=0.9\columnwidth]{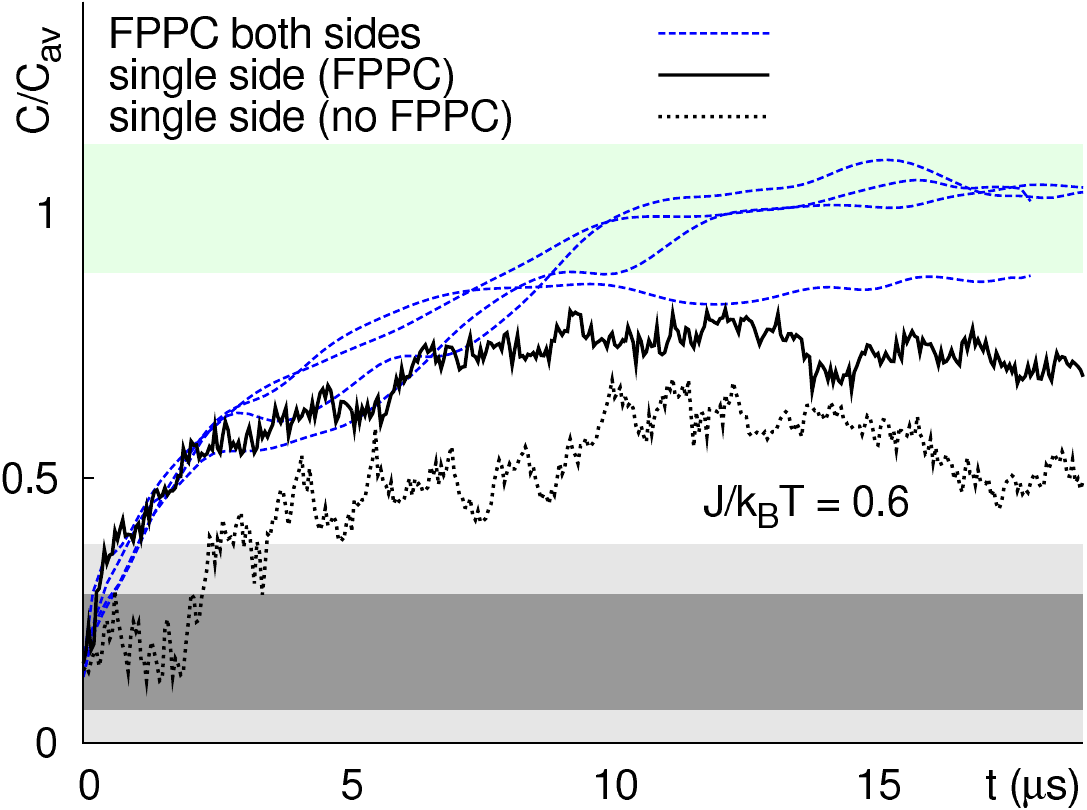}
\end{center}

\caption{ \label{fig:Aalign} Compatibility of our MD simulations with the 
average domain structure of the Ising model in \fig{fig:Avalues}. The 
compatibility is normalized by $C_{\rm av}$, which is the average compatibility 
of the Ising model with its own average structure. The upper shaded region 
corresponds to one standard deviation around this average. The inner and outer 
lower shaded areas represent one standard deviation regions that are expected 
for randomly located domains (corresponding to $r=2$~nm and $r=7.5$~nm, 
respectively; see details in text). The MD simulations with FPPC obstacles in 
both leaflets (runs IIa and IIb) equilibrate to states consistent with the Ising 
model (i.e.~around the upper shaded region) and are clearly incompatible with 
random domains. In the case of FPPC in only one leaflet (run~III), the 
$C$-values are still distinctly different from random domains (i.e.~outside the 
lower shaded region), but neither the FPPC leaflet (solid curve) nor the 
\ahum{free leaflet} (dashed curve) appear to be fully compatible with the Ising 
model.}

\end{figure}


Since the $A_i$ will in general have values {\it between} $-1$ and $+1$, and 
since typical snapshots [cf.~\fig{fig:IsingStates}(b,c,d)] do not exactly have 
the domain structure of the average snapshot [\fig{fig:Avalues}], we do not 
expect the compatibilities of our MD simulation to reach $C \approx 1$. Instead, 
the proper $C$ values to compare against are those of the Ising model snapshots 
themselves, which are straightforward to obtain: We simply let the MC simulation 
that was used to produce \fig{fig:Avalues} run a second time and collect a 
histogram of $C$ values. The histogram is Gaussian, with mean $C_{\rm av}$ and 
width $\sigma$; the upper shaded region in \fig{fig:Aalign} marks the interval 
$(C_{\rm av} \pm \sigma)/C_{\rm av}$. In case the domains obtained in the MD 
simulations are compatible with the averaged Ising structure of 
\fig{fig:Avalues}, we expect the corresponding compatibilities to be inside or 
at least in the proximity of the upper shaded area. Indeed, as \fig{fig:Aalign} 
shows, the MD compatibilities profoundly increase with time, indicating that the 
corresponding domain structures increasingly correlate with the equilibrium 
domains of the Ising simulation. As may be expected, the correlation is most 
pronounced in the case where the FPPC obstacles reside in both leaflets, where 
an excellent agreement with the Ising model is found. However, also when FPPC 
resides in only one leaflet, the correlation remains. In this case, the 
correlation is most pronounced in the leaflet containing the obstacles, as might 
be expected.


To demonstrate that the agreement between the domain structures of the Ising 
model and the MD simulations does not merely occur by chance, we also measure 
the typical compatibilities that one obtains in case the domains appear at 
random locations. To generate a random domain structure, we take the initial 
state of the MD simulation, but with all phospholipids (except FPPC) being 
DLiPC. Then, we pick a random location in the membrane and change the identity 
of all DLiPC lipids within a radius $r$ of this location to DPPC. This step is 
repeated until the (DPPC:DLiPC) ratio equals that of the simulation (note that 
the resulting domains are usually not circular, since the circular regions 
picked at each step can overlap). Since this process is fast, we can generate 
many such random domain structures and collect a histogram of compatibilities. 
This histogram is also Gaussian, but centered around a lower average value, and 
with a different width that depends on~$r$. The range of compatibilities 
expected for these domain structures is marked in \fig{fig:Aalign} by the lower 
shaded areas. As can be seen in the figure, they are clearly distinct from the 
MD curves (which holds for all values of $r$). We therefore conclude that the 
correspondence between domains obtained in our MD simulations and the Ising 
model is not coincidental. Despite the very different interaction potentials of 
the two models, both yield equivalent domain structures. Consequently, we expect 
that the domains seen in our MD simulations are indeed equilibrated states, as 
opposed to kinetically trapped states that would eventually macroscopically 
phase separate.


We emphasize that our findings are not unique to $J/ \kT =0.6$ considered here. 
We have also performed the analysis for $J / \kT =0.45,0.5,\dots ,0.65$, and 
always found the same scenario: A reasonable agreement between the MD 
simulations with obstacles on both sides and the Ising model, while the 
\ahum{pure chance} regime is always excluded. The reason for this agreement is 
that over a wide range of temperature, the average domain structure 
[\fig{fig:Avalues}] qualitatively remains the same, and merely becomes sharper 
as the temperature is decreased~\cite{citeulike:8864903, Machta20111668}.

\section{Summary \label{sec:summary}}

In this work, we have presented results from MD simulations of a 
DPPC/DLiPC/cholesterol membrane at near-atomic resolution in the absence and in 
the presence of quenched disorder. Our results show that static components in 
the membrane destroy the formation of large domains. Instead, a heterogeneous 
structure of small domains is seen [\fig{fig:obstacles}]. For this to happen, it 
suffices that the obstacles are present in only one of the leaflets, since 
quenched disorder effects can be transmitted to the other leaflet via 
inter-leaflet coupling [\fig{fig:single}]. As the results have been obtained for 
a system with a strong demixing tendency, the elimination of macroscopic domain 
formation should occur for other lipid compositions also. A coupling of a cell 
to its surroundings or its cytoskeleton that causes an immobilization of 
membrane components may therefore explain the appearance of raft-like domains in 
thermal equilibrium.

By means of a suitably constructed compatibility measure we have also shown that 
when the obstacles reside in both leaflets, the domain structure can be 
predicted with very simple models, such as the Ising model 
(cf.~\fig{fig:Aalign}). We do not expect this to be a feature unique to the 
Ising model, but instead consider this result a signal that structure formation 
in the presence of quenched disorder is universal and does not strongly depend 
on the microscopic details of the system. Other similarly simple membrane 
models~\cite{citeulike:9234255, citeulike:8864903, pinkmodel} are expected to be 
equally suited to predict domain structure. The exception is in cases where the 
obstacle configurations differ between the leaflets. In this situation, a simple 
monolayer model may be inadequate due to the neglect of inter-leaflet coupling. 
The construction and use of a related bilayer model is conceivable but has not 
been tested.

If one accepts the evidence that a MC simulation of a simple model can 
adequately describe the equilibrium domain structure of membranes in the 
presence of quenched disorder, their comparably low computational demands allow 
considering questions that are inaccessible in detailed MD simulations. The 
possibility to probe larger length scales permits the investigation of obstacle 
configurations with a structure on their own~\cite{Ehrig201180, Machta20111668}, 
and connect to the length scales of experimental visualization 
methods~\cite{Hell:STED, Eggeling:STED}, while the option to simulate many 
different setups allows to compare different positionings and different types of 
static obstacles\cite{citeulike:8864903,citeulike:7115548}.

\subsection*{Acknowledgments}

This work was supported by the {\it Deutsche Forschungsgemeinschaft} via the 
Emmy Noether program~(VI~483) and the SFB~803~(project B2).

\bibliography{references}

\end{document}